\documentclass[twocolumn,showpacs,preprintnumbers,amsmath,amssymb]{revtex4-1}
\usepackage{tabularx,graphicx}
\graphicspath{{figures/}} 
\usepackage{hyperref}
\hypersetup{
    colorlinks=true,
    linkcolor=blue,
    filecolor=magenta,      
    urlcolor=blue,
}

\usepackage{comment}
\usepackage{empheq}
\def\comment#1{}

\usepackage{color}

\begin{document}

\title{Asymmetric wave functions from tiny perturbations}
\author{T.~Dauphinee  and F.~Marsiglio}
\email{tdauphin@ualberta.ca, fm3@ualberta.ca}
\affiliation{Department of Physics, University of Alberta, Edmonton, AB, Canada T6G 2E1}

\begin{abstract}
\noindent The quantum mechanical behavior of a particle in a double well defies our intuition based on classical reasoning.
Not surprisingly, an asymmetry in the double well will restore results more consistent with the classical picture. What is surprising,
however, is how  a very small asymmetry can lead to essentially classical behavior. In this paper we use the simplest version
of a double well potential to demonstrate these statements. We also show how this system accurately maps onto a two-state system, which we refer to as a `toy model'. 
\end{abstract}

\date{\today} 
\maketitle

\section{Where is the particle?}

Given two wave functions, the one shown in Fig.~\ref{fig1}(a) and the one shown in Fig.~\ref{fig1}(b), which is the correct eigenstate for a double well potential?

The actual correct answer is that we do not have enough information. We do not have enough information because we are only given a picture of the double well potential, and not an explicit definition. An explicit definition of the potential used in Fig.~\ref{fig1} will be provided below, and this definition will reveal an asymmetry not apparent in the figure, that the potential well on the right hand side is slightly lower than that on the left hand side. The potential is actually drawn with this asymmetry, but it is so small as to be invisible to the eye on this scale (by many orders of magnitude); 
the net result, however, is that the actual ground state is that pictured in 
Fig.~\ref{fig1}(b), with the wave function located entirely in the right side well. Hence, a tiny perturbation (the `flea')
results in a state very different from the familiar symmetric superposition of `left' and `right' well occupancy (shown in Fig.~\ref{fig1}(a)).
\par
\bigskip

\section{The Asymmetric Double Well Potential}

\subsection{Introduction}

\begin{figure}[here]
\includegraphics[width =8.0cm,angle=0]{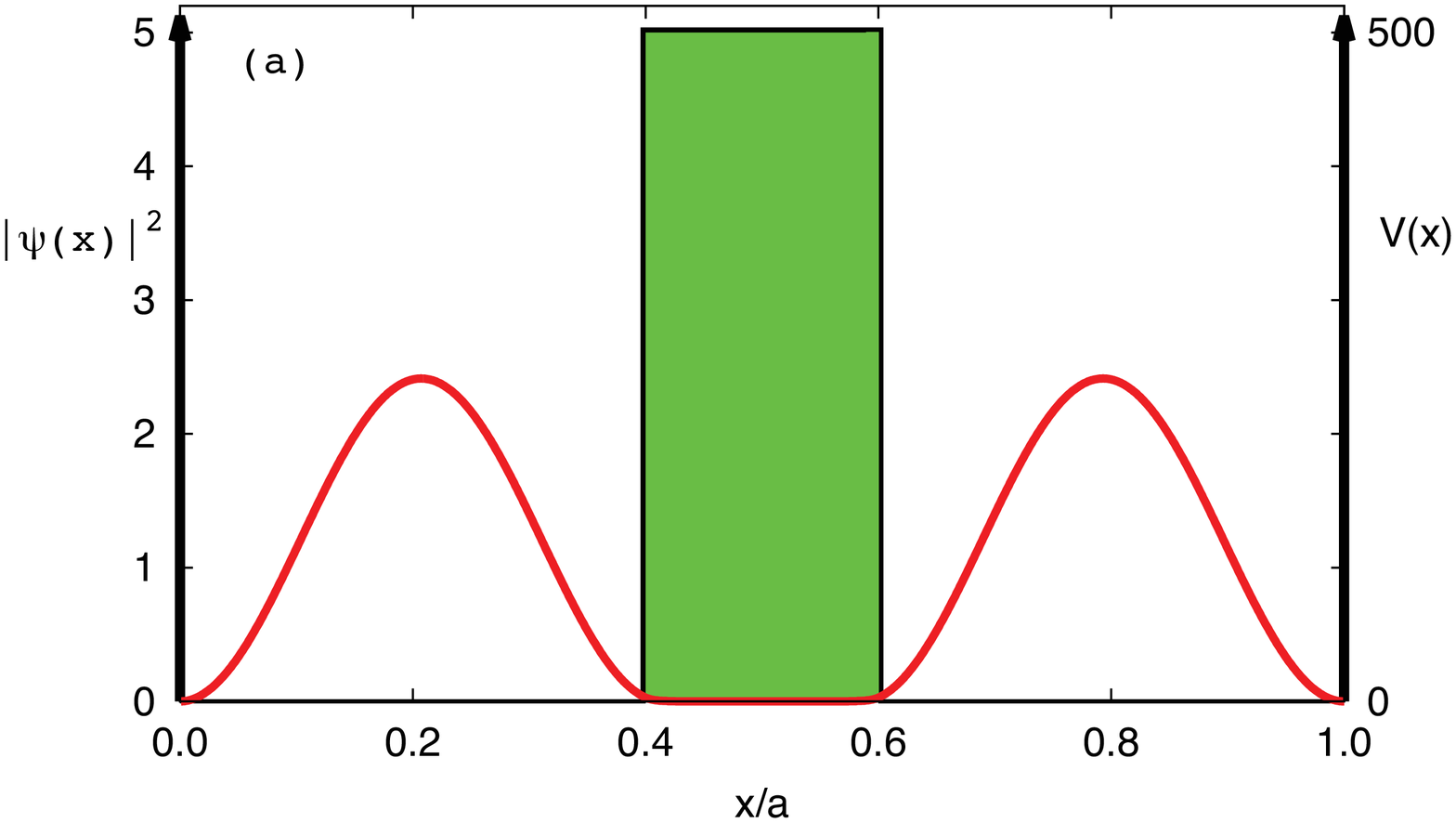}
\includegraphics[width =8.0cm,angle=0]{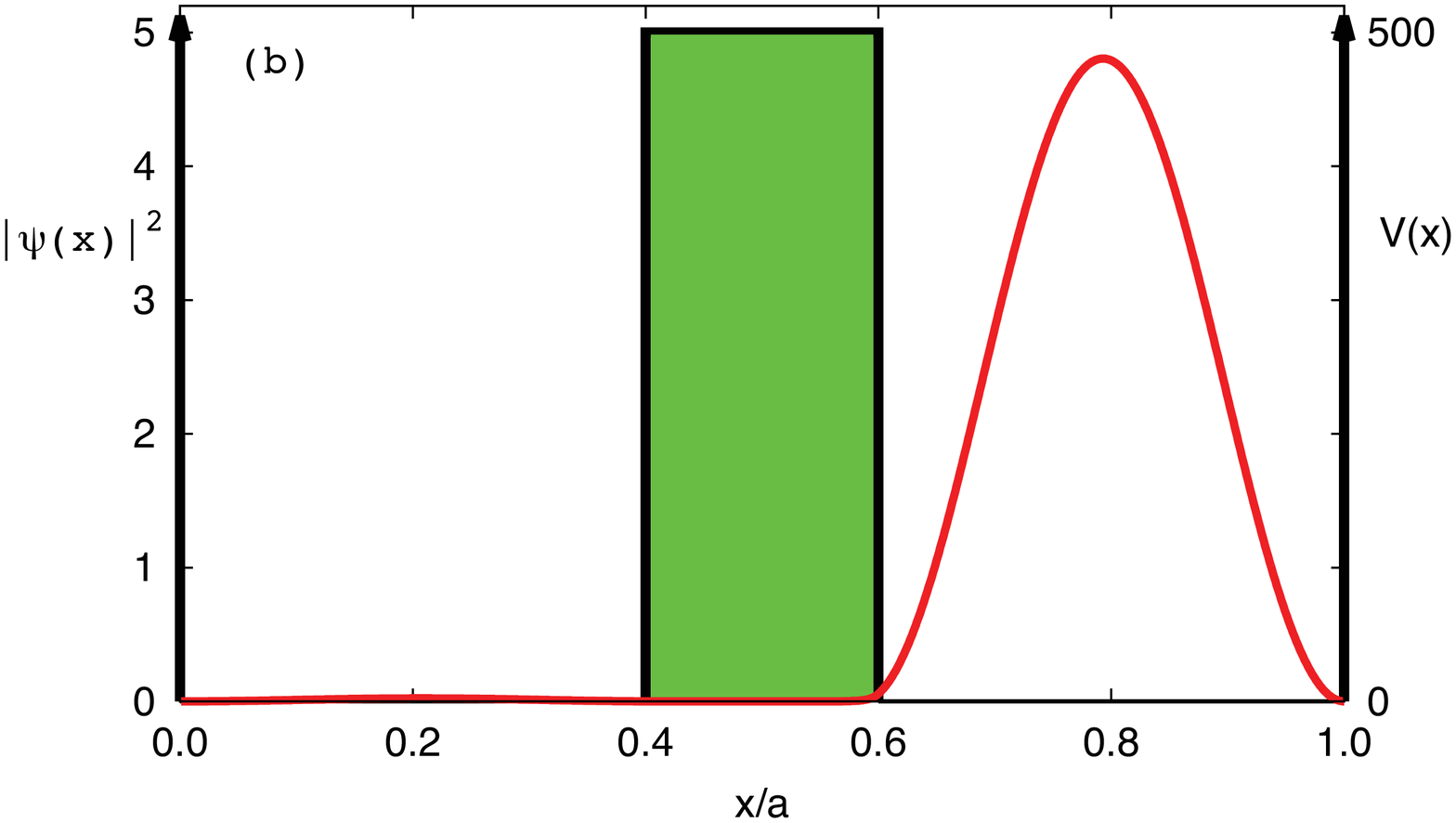}
\caption{(a) State A, ``Left'' \textbf{and} ``Right'', or (b) State B: ``Just Left'' \textbf{or} ``Just Right''. In the first state the wave function
has equal amplitude to be in either well, a linear superposition of states so common in quantum mechanics, while in the second state the wave function
is entirely in the right side well. Which is the ground state for the double well potential shown?}
\label{fig1}
\end{figure}

The double well potential is often used in quantum mechanics to illustrate \hypertarget{hrazavy03}{situations} \hypertarget{hjelic12}{in} which more than one state is accessible
in a system, with a coupling from one to the other through tunnelling.\cite{razavy03,jelic12} For example, in the \hypertarget{hfeynman65}{Feynman Lectures} the
ammonia molecule is used to illustrate a physical system that has a double well potential for the Nitrogen atom.\cite{feynman65}
He used an effective two-state model to illustrate these ideas, and in most undergraduate textbooks a two state system is utilized
for a similar purpose. Here instead we will first focus on a full solution to a double well potential; the features inherent in a two state
system will emerge from our calculations. Indeed, we will also present a refined two-state model to
capture the essence of the asymmetry in a microscopic double well potential, as we are using in Fig.~\ref{fig1}.

The notion that a slight asymmetry can result in a  \hypertarget{hjona-lasinio81a}{drastic} \hypertarget{hjona-lasinio81b}{change} \hypertarget{hsimon85}{in} 
the wave function is not new --- it was
first discussed in Refs. [\onlinecite{jona-lasinio81a,jona-lasinio81b,simon85}], but in a manner and context inaccessible to
\hypertarget{hreuvers12}{undergraduate} \hypertarget{hlandsman12}{students}. More recently the topic has been revisited \cite{reuvers12,landsman12} to illustrate an emerging
phenomenon in the semiclassical limit ($\hbar \rightarrow 0$). These authors refer to the `Flea' in reference to
the very minor
perturbation in the potential (as in ours) and to the `Elephant' \cite{simon85} (the very deep double well potential). 
Schr\"odinger's Cat has crept into the discussion \cite{reuvers12,landsman12} because the `flea' disrupts the entangled
character (Fig.~\ref{fig1}(a)) of the usual Schr\"odinger cat-like double well wave function (Fig.~\ref{fig1}(a)).

The purpose of this paper is to utilize a very simple model of an asymmetric double well potential, \hypertarget{hmarsiglio09}{solvable} either analytically 
or through an application of matrix mechanics, \cite{marsiglio09,jelic12} to demonstrate the rather potent effect of a rather tiny
imperfection in the otherwise symmetric double well potential. Contrary to the impression one might get from the references on this subject, there
is nothing `semiclassical' about the asymmetry of the wave function illustrated in Fig.~\ref{fig1}(b). We will show,
using a slight modification of Feynman's ammonia example, \cite{feynman65,landsman12} that the important parameter to
which the asymmetry should be compared is the tunnelling probability; this latter parameter can be arbitrarily small. This correspondence applies for excited states as well, along with other asymmetric double well shapes.

\subsection{Square Double Well with Asymmetry}

\begin{figure}[here]
\includegraphics[width=0.55\textwidth]{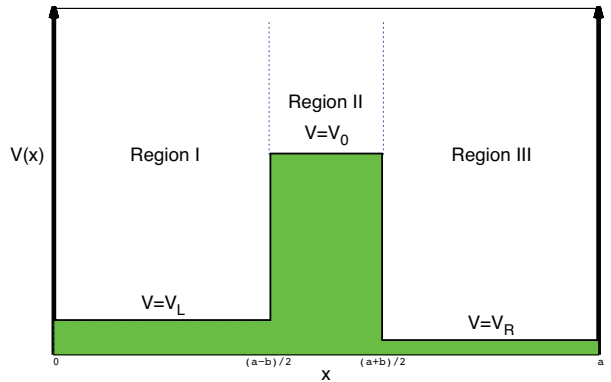}
\caption{A schematic of the generic asymmetric square double well potential. The well widths are the same but the
left- and right-side levels can be independently adjusted.}
\label{fig2}
\end{figure}

A variety of symmetric double well potentials was used in Ref. \onlinecite{jelic12} to illustrate the universality of the
energy splitting described there. Here instead we will use perhaps the simplest model to exhibit the impact of asymmetry --- we have checked with other versions and the same physics applies universally. This model uses two square wells, with left and
right wells having base levels $V_L$ and $V_R$, respectively, separated by a barrier of width $b$ and height $V_0$
and enclosed within an infinite square well extending from $x=0$ to $x=a$. For our purposes we assign the two wells
equal width, $w = (a-b)/2$. Mathematically it is described by
\begin{equation}
V(x) = \begin{cases} \infty & \text{if $x < 0$ or $x > a$} \\ 
V_0 & \text{if $(a-b)/2 <  x < (a+b)/2$} \\ 
V_L & \text{if $0 < x < (a-b)/2$} \\ 
V_R & \text{if $(a+b)/2 < x < a$},
\end{cases}
\label{asy_doublewell_potential} 
\end{equation}
and is shown in Fig.~\ref{fig2}. We can readily recover the symmetric double well by using $V_L = V_R$. Units of 
energy are those of the ground state for the infinite square well of width $a$, $E_1^0 = \pi^2 \hbar^2/(2m_0a^2)$, 
where $m_0$ is mass of the particle, and units of length are expressed in terms of the width of the infinite square well, $a$.
We will typically use a barrier width $b = a/5$ so that the individual wells have widths $w = 2a/5$. The height of the barrier
then controls the degree to which tunnelling from one well to the other occurs, and $\delta \equiv (V_L - V_R)/2$ controls
the asymmetry.

In the following subsection we will work through detailed solutions to this problem, first analytically, and then numerically.
It is important to realize that these solutions retain the full Hilbert space in the problem. A `toy' model is introduced in a later
subsection, and reduces this complex problem to a two-state problem. We then proceed to illustrate how the two-state problem
reproduces remarkable features of the complex problem, as a function of the asymmetry in the two wells.

\subsection{Preliminary Analysis}

Fig.~\ref{fig1} was produced with $v_0 \equiv V_0/E_1^0 = 500$, $b/a = 0.2$, $v_L \equiv V_L/E_1^0 = 0$, and
$v_R \equiv V_R/E_1^0 = 0$ in part (a) and $v_R \equiv V_R/E_1^0 = -0.00001$ in part (b). The change in potential strength
compared with the barrier between these two cases is $1$ part in $50,000,000$. This results in ground state energies of
$e_1 \equiv E_1/E_1^0 = 5.827 034$ and $e_1 \equiv E_1/E_1^0 = 5.827 025$ for the symmetric and asymmetric case,
respectively. Needless to say, either the difference in potentials or the difference in ground state energies represent minute
changes compared to the tremendous qualitative change in the wave function evident in parts (a) vs (b) of Fig.~\ref{fig1}.

The potential used is simple enough that an `analytical' solution is also possible. The word `analytical' is in quotations here
because, in reality, the solution to the equation for the energy for each state must be obtained graphically, i.e. numerically. 
While this poses no significant difficulty,
it is sufficient \hypertarget{hremark1}{work} that essentially all textbooks stop here, and do not examine the wave function.\cite{remark1} 

Assuming that $E<V_0$, the analytical solution is
\begin{align}
\psi_{I}(x) & = A\sin{kx} &  k  = \left( 2m(E - V_L)/\hbar^{2}\right)^{\frac{1}{2}} \notag \\
\psi_{II}(x) & = Be^{\kappa x} + Ce^{-\kappa x} & \kappa  = \left( 2m(V_{0}-E)/\hbar^{2} \right)^{\frac{1}{2}} \notag \\
\psi_{III}(x) & = D\sin{q(a-x)} & q = \left( 2m(E - V_R)/\hbar^{2}\right)^{\frac{1}{2}}.
\label{psi_regions}
\end{align}
where the regions I, II, and III are depicted in Fig.~\ref{fig2}.
Applying the matching conditions at $x = (a \pm b)/2$ leads to an equation for the allowed energies: 
\begin{align}
  \left( \sin(kw) +  \frac{k}{\kappa}\cos(kw) \right) & \left( \sin(q w) + \frac{q}{\kappa}\cos(q w)\right) = \notag \\
  e^{-2 \kappa b} \left( \sin(kw) - \frac{k}{\kappa}\cos(kw)  \right) & \left( \sin(q w) - \frac{q}{\kappa}\cos(q w) \right),
\label{allowed}
\end{align}
where $w \equiv (a-b)/2$ is the width of the individual wells and $b$ is the width of the barrier. The matching conditions also
provide expressions for the relative amplitude $D/A$:
\begin{equation}
\left| \frac{D}{A} \right| =  e^{-\kappa b}  \left| \frac{\kappa \sin(kw) - k\cos(kw) }{\kappa \sin(qw) + q\cos(qw)} \right|,
\label{relamp}
\end{equation}
or, alternatively,
\begin{equation}
\left| \frac{D}{A} \right| =  e^{+\kappa b} \left| \frac{\kappa \sin(kw) + k\cos(kw) }{\kappa \sin(qw) - q\cos(qw)} \right|.
\label{relampinv}
\end{equation}
To proceed further, Eq.~(\ref{allowed}) is solved numerically for the allowed energy values. 
As might be expected the low energy solutions come in pairs. 
Once $E$ is known,
then so are $k$, $q$ and $\kappa$, and the $D/A$ ratio can be determined through either of Eqs.~(\ref{relamp}) or (\ref{relampinv}).
In addition, through similar relations and normalization, all other coefficients in Eq.~(\ref{psi_regions}) can be determined.
Notice that at first glance Eq.~(\ref{relamp}) suggests that $|D| << |A|$, while  Eq.~(\ref{relampinv}) suggests the opposite.
In reality both equations provide the correct answer, though with limited numerical precision one is generally more accurate than the other. 
Which is more accurate depends on whether
$V_R < V_L$ or vice-versa. For $V_R {{ \atop <} \atop {>\atop }} V_L$, we expect $|D| {{ \atop >} \atop {<\atop }} |A|$.

Alternatively, we solve the original Schr\"odinger Equation numerically right from the start
by expanding in the infinite square well basis [$\phi_n(x)
= \sqrt{2 \over a} \sin{\left({n \pi x \over a}\right)}$ for $n = 1,2,3....$], and `embed' the double well part in this basis.
More specifically, we write
\begin{equation}
|\psi\rangle = \sum_n c_n |\phi_n\rangle,
\label{expansion}
\end{equation}
and insert this into the Schr\"odinger Equation to obtain the matrix equation,
\begin{equation}
\sum_m H_{nm} c_m = E_n c_n,
\label{matrix}
\end{equation}
where
\begin{eqnarray}
&&H_{nm} = \delta_{nm} \left( n^2E_1^{0} +(V_L+V_R)w/a + V_0 b/a \right) \nonumber \\
&& \phantom{aaaaa}+\delta_{nm}  [2V_0-V_L-V_R] \ {\rm sinc}(2n) \nonumber \\
&&+\left(1 - \delta_{nm} \right) D_{nm} \left(V_L - V_0 + [V_R - V_0] (-1)^{n+m}]\right) \nonumber \\
\label{ham}
\end{eqnarray}
where
\begin{equation}
D_{nm} \equiv  {\rm sinc}(n-m)  - {\rm sinc}(n+m),
\label{defn}
\end{equation}
and
\begin{equation}
{\rm sinc}(n) \equiv {{\rm sin}(\pi n w) \over \pi n}.
\label{sinc}
\end{equation}
As before,   $w \equiv (a-b)/2$ and $b$ are the widths of the wells and barrier, respectively.
The general procedure is provided in Refs. [\onlinecite{marsiglio09}] and [\onlinecite{jelic12}], and the reader is referred 
to these papers for more details. Using either the analytical expressions or the numerical diagonalization, the results are identical. The advantage of the latter method is that the study is not confined to simple well geometries consisting of boxes, and
students can easily explore a variety of double well potential shapes.

\subsection{Results}
In Fig.~3 the resulting wave function is shown for a variety of asymmetries.
\begin{figure}[here]
\includegraphics[width =0.5\textwidth]{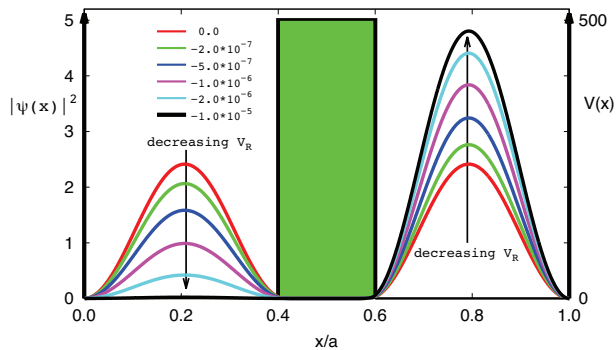}
\caption{Progression of the wave function as the right well level is lowered from $0$ (same as the left well level) to
$-10^{-5}$ (in units of $E_1^0 = \hbar^2 \pi^2/(2ma^2)$). When the double well is symmetric the probability 
density ($\left|\psi(x)\right|^2$) is symmetric (shown in red); the degree of asymmetry in the probability density increases
monotonically as $V_R/E_1^0$ decreases from $0$ to $-10^{-5}$. The actual values of $V_R$ used in this plot, in units of 
$E_1^0$, are indicated in the legend. Any of these values is absolutely
indistinguishable on the energy scale of the barrier ($V_0/E_1^0 = 500$) or the ground state energies ($E_{\rm GS}/E_1^0
\approx 5.827$).
}
\label{fig3}
\end{figure}
The result is remarkable. As $V_R/E_1^0$ decreases from $0$ to a value of $-0.00001$, the probability density changes from a
symmetric profile (equal probability in left and right wells) to an entirely asymmetric profile (entire probability localized
to the right well). The other `obvious' energy scales in the problem are the barrier height ($V_0/E_1^0 = 500$)  and the
ground state energies ($E_{\rm GS}/E_1^0 \approx 5.827$), so these changes in the potential are minute in comparison. 
Even more remarkable 
is that as far as the energies are concerned, these minute changes give rise to equally minute changes in the ground state
energy: $E_{\rm GS}/E_1^0 \approx 5.827034$) ($ \approx 5.827025$) for $V_R/E_1^0 = 0.0$ ($V_R/E_1^0 = -0.000 01$),
respectively, while the changes in the wave functions are qualitatively spectacular. 

\subsection{Discussion}

That such enormous qualitative changes can result from such minute asymmetries in the double well potential is of
course important for experiments in this area, where it would be very difficult to control deviations from perfect symmetry in a typical double well potential
at the $10^{-5}$ level. Why is this phenomenon not widely disseminated in textbooks? And what precisely controls the
energy scale for the `flea-like' perturbation that eventually results in a completely asymmetric wave function situated in only one
of the two wells [Fig.~(1b)]? The answer to the first question is undoubtedly connected to the lack of a straightforward analytical
demonstration of the strong asymmetry in the wave function. Certainly Eqs. (\ref{relamp},\ref{relampinv}) exist, but it is difficult to coax out of 
either of these equations an explicit demonstration of the resulting asymmetry apparent in Fig.~(3) as a 
function of lowering (or raising) the level of the potential well on the right. To shed more light on this phenomenon, 
and to provide an answer to the second question, we resort to a `toy model' slightly modified from the one used by 
Feynman \cite{feynman65} to explain tunnelling in a symmetric double well system, and introduced more 
recently by Landsman and Reuvers \cite{landsman12,reuvers12} in a perturbative way. Such a model is also \hypertarget{hcohentannoudji77}{used} 
in standard  \hypertarget{htownsend00}{textbooks}
to discuss `fictitious' spins interacting with a magnetic field\cite{cohentannoudji77} and two level systems subject to an electric field.\cite{townsend00}

\section{A Toy model for the Asymmetric Double Well}

An aid towards understanding the results of our `microscopic' calculations is provided by an `effective' model. The tact is to
strip the system of its complexity and focus on the essential ingredients. In this instance, the key features amount to whether the
particle is in the right well, or left well, or a combination thereof.
Following Feynman, we begin with two isolated wells, each with a particular energy level, and with each coupled to the other
through some matrix element, $t$:
\begin{eqnarray}
H\psi_L &=& E_L \psi_L - t\psi_R \nonumber \\
H\psi_R &=& E_R \psi_R - t\psi_L,
\label{feynman}
\end{eqnarray}
where, in the absence of coupling, the left (right) well would have a ground state energy $E_L$ ($E_R$), and $\psi_L$
($\psi_R$) represents a wave function localized in the left-side (right-side) well. A straightforward solution of this two state
system results in an energy splitting, as in the symmetric case:
\begin{equation}
E_{\pm} = {E_L + E_R \over 2} \pm \sqrt{\left({E_L - E_R \over 2}\right)^2 + t^2}.
\label{energy_split}
\end{equation}
For typical barriers ($t <<(E_L + E_R)/2$) and small asymmetries ($E_L - E_R << (E_L + E_R)/2$), very little difference occurs in the energies, in agreement with the results from our more microscopic calculations above.
If we define $\delta \equiv (E_L - E_R)/2$ [$ \equiv (V_L - V_R)/2$ for the square double well potential], then the 
ground state wave function becomes
\begin{equation}
\psi = {1 \over \sqrt{2}} \sqrt{1 - {\delta \over \sqrt{\delta^2 + t^2}}}\psi_L + {1 \over \sqrt{2}} 
\sqrt{1 + {\delta \over \sqrt{\delta^2 + t^2}}}\psi_R.
\label{wave_function_asym}
\end{equation}
In the symmetric case we recover the (symmetric) linear superposition of the state with the particle in the left well, along
with the state with the particle in the right well (see the remark in Ref. [\onlinecite{remark1}]). With increasing 
asymmetry, however, say with $V_R < V_L$, i.e. $\delta > 0$,
the amplitude for the particle being in the right well rises to unity, while that for the particle in the left well decreases to zero.
Our toy model illustrates that the energy scale for this cross-over is the tunnelling matrix element, $t$. This energy scale must be
clearly present in the microscopic model defined in Eq.~(\ref{asy_doublewell_potential}), but it is not there explicitly.

\subsection{Comparison of the toy model to the microscopic model}

To see how well the toy model defined by the two-state system in Eq.~(\ref{feynman}) reproduces properties of the
microscopic calculations, we make an attempt to
compare the results from the two calculations. This is most readily accomplished by the following procedure. First,
the solid curves displayed in Fig.~4 are readily obtained by plotting the two amplitudes in Eq.~(\ref{wave_function_asym}), 
\begin{eqnarray}
|c_L|^2 &\equiv &{1 \over 2} \left( 1 - {\delta \over \sqrt{\delta^2 + t^2}} \right) \nonumber \\
|c_R|^2 &\equiv &{1 \over 2} \left( 1 + {\delta \over \sqrt{\delta^2 + t^2}} \right)
\label{cs}
\end{eqnarray}
as a function of $\delta/t$. Then, for one of the results shown in Fig.~(\ref{fig3}),
\begin{figure}[here]
\includegraphics[width =0.56\textwidth]{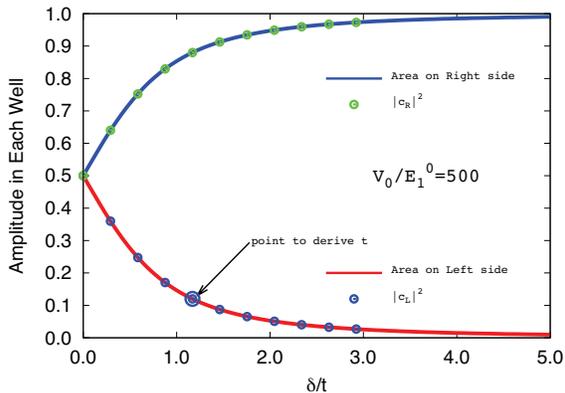}
\caption{Plot of the amplitude in each well vs the asymmetry parameter, $\delta/t$. The circles indicate the area integrated from Fig.~3 on each side of the barrier, as indicated. The large circle shows the point used to establish a value of $t \approx 6.84\cdot 10^{-7} E_1^0$, so that the expression in Eq.~(\ref{cs}) matches precisely the value determined by the more microscopic calculation of the previous subsection. With this value of $t$ the curves corresponding to the expressions in Eq.~(\ref{cs}) are also plotted, and the agreement is excellent over the entire range of $\delta/t$. This indicates that the `toy model' phenomenology 
is very accurate. }
\label{fig4}
\end{figure}
we compute the area under the curve on the left; it will correspond to an amplitude $|c_L|^2$ in Fig.~4. By placing
this value on the appropriate curve in Fig.~4 we are able to extract a value of $\delta/t$ and hence an effective value
of $t$ (since $\delta \equiv (V_L- V_R)/2$ is known). This is marked by a large circle in Fig.~4.
We have thus identified a value of $t$, strictly only defined for the
toy model, with a specific barrier height and width in the more microscopic calculations connected with Eqs. (\ref{psi_regions}-
\ref{relampinv}) or their numerical counterparts. We can then vary the value of $\delta$ (as was done to generate the 
curves shown in Fig.~3) and plot the values of the total probability density in the left and right wells as a function of $\delta/t$. 
The smaller circles in Fig.~4 are the results of these calculations, and they almost perfectly lie on the curves generated
from the toy model, thus showing that the asymmetric double well system indeed behaves like a two-state system
described phenomenologically by Eq.~(\ref{feynman}). We have done this for other barrier heights and widths and similar
very accurate agreement between the two approaches is achieved. 
We have also carried out such comparisons for excited states, and
also for other kinds of double wells (e.g. so-called Gaussian wells), with similar agreement.

\begin{figure}[here]
\includegraphics[width =0.5\textwidth]{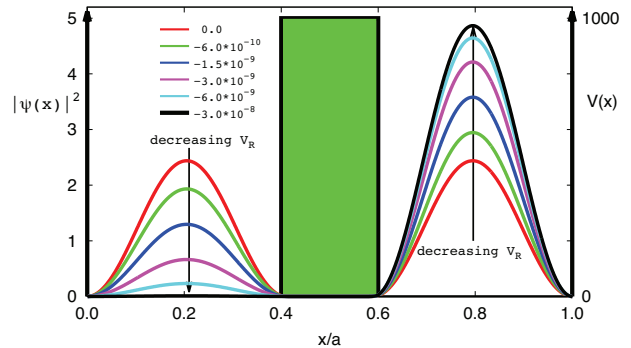}
\caption{As in Fig.~3, progression of the wave function as the right well level is lowered from $0$ (same as the left well level) to
$-3.0\cdot 10^{-8}$ (in units of $E_1^0 = \hbar^2 \pi^2/(2ma^2)$), but now with a barrier height of $V_0 = 1000E_1^0$. When the double well is symmetric the probability 
density ($\left|\psi(x)\right|^2$) is symmetric (shown in red); the degree of asymmetry in the probability density increases
monotonically as $V_R/E_1^0$ decreases from $0$ to $-3.0\cdot 10^{-8}$. The actual values of $V_R$ used in this plot, in units of 
$E_1^0$, are indicated in the legend. Even more so than before, any of these values is absolutely
indistinguishable on the energy scale of the barrier ($V_0/E_1^0 = 1000$) or the ground state energies ($E_{\rm GS}/E_1^0
\approx 5.947$).}
\label{fig5}
\end{figure}

\begin{figure}[here]
\includegraphics[width =0.56\textwidth]{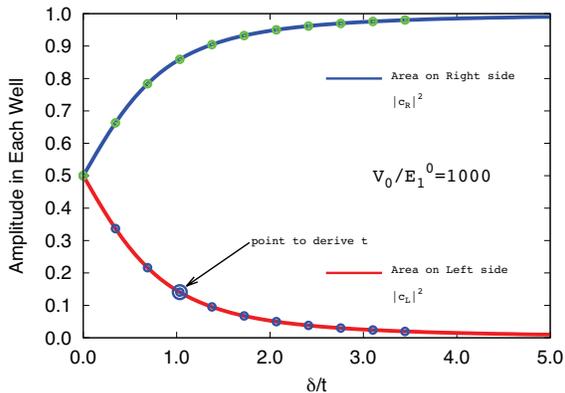}
\caption{As in Fig.~4, plot of the amplitude in each well vs the asymmetry parameter, $\delta/t$, but now for a much higher barrier potential, $V_0/E_1^0 = 1000$. The circles indicate the area integrated from Fig.~5 on each side of the barrier, as indicated. The large circle shows the point used to establish a value of $t \approx 1.45\cdot 10^{-9} E_1^0$, so that the expression in Eq.~(\ref{cs}) matches precisely the value determined by the more microscopic calculation of the previous subsection. With this value of $t$ the curves corresponding to the expressions in Eq.~(\ref{cs}) are also plotted, and the agreement is excellent over the entire range 
of $\delta/t$. As long as the barrier is sufficiently high to delineate two very distinct states (`left' and `right,' or `dead' and `alive',
the two state model works very well.}
\label{fig6}
\end{figure}

As an example, in Fig.~5 and Fig.~6 we show results analogous to those of Fig.~3 and Fig.~4, but for a double well with a barrier
with the same width, but with a significantly increased height. The sequence of probability densities in Fig.~5 is similar
to those shown in Fig.~3 except that the changes in the potential asymmetry are orders of magnitude smaller. Fig.~6 then
confirms that the significantly enhanced sensitivity is due to the significantly reduced effective `hopping' amplitude $t$
between the two wells, such that the asymmetry in probability densities as a function of potential asymmetry in Fig.~6
is as it is in Fig.~4 as a function of $\delta/t$, with both of these parameters greatly reduced.

\subsection{Origin of the coupling $t$}

The origin of the coupling parameter $t$ in the two-state  toy model is clearly the possibility of tunnelling that exists from one well into
the other. It is rather involved to `derive' this parameter $t$ from parameters of the original double well potential 
\hypertarget{hmerzbacher98}{specified} in Eq.~(\ref{asy_doublewell_potential}). 
In fact it suffices to provide an estimate based only on the symmetric case ($V_L = V_R = 0$), 
and we provide a brief exposition here, following Merzbacher.\cite{merzbacher98}

One starts with a variational wave function, of the form 
\begin{equation}
\psi_{\pm}(x) = {N_{\pm} \over \sqrt{2}}\bigl( \psi_L(x) \pm \psi_R(x) \bigr),
\label{variational}
\end{equation}
where the $\pm$ refers to the ground state ($+$) or first excited state ($-$), respectively, and the subscript $L$ ($R$)
refers to the left (right) well, respectively. First taking the two wells in isolation, we obtain $\psi_L(x)$, for example, with a solution
similar to that in Eq.~(\ref{psi_regions}),
\begin{align}
 \psi_L(x) & = A\sin{(kx)} & 0<x < w \notag \\
 & = A\sin{(kw)} e^{-\kappa(x-w)}  & x > w,
\label{psi_L_region}
\end{align}
and similarly for the well on the right; with all coordinates displaced a distance $b$ to the right, it forms a mirror image of the one on the left. 
This distance can be considered
to be very large at first. In these equations we have included an unimportant normalization constant, $N_{\pm}$ in Eq.~(\ref{variational})
and $A$ in Eq.~(\ref{psi_L_region}). The energy splitting between the two states is determined by the `overlap' between the two
wells. A straightforward calculation gives
\begin{equation}
E_{\rm split} = \int \ dx \ \psi_L(x) H \psi_R(x) \propto e^{-\kappa b} \approx e^{-b\sqrt{V_0}},
\label{split}
\end{equation}
so as $V_0$, the height of the barrier, increases for fixed width $b$, the splitting becomes less and less. Note, however, that the
parameter $t$, first introduced in Eq.~(\ref{feynman}) is proportional to this same quantity:
\begin{equation}
t \propto e^{-b\sqrt{V_0}}.
\label{t}
\end{equation}
This factor is equal to $8.6 \times 10^{-7}$ and $2.5 \times 10^{-9}$ for $V_0/E_1^0 = 500$ and $1000$, respectively. The actual values
of $t$ obtained phenomenologically through the fitting procedure described above are $6.8 \times 10^{-7}$ and $1.5 \times 10^{-9}$,
respectively, which tracks very closely these exponentially decaying factors.

\section{Summary}

We have examined the simplest asymmetric double well potential and explored the behavior of the wave function
as a function of asymmetry. In the symmetric case the ground state is a linear superposition of the particle in the left
well and the particle in the right well.
As the floor level of the potential on the right side ($V_R$) decreases, the probability
for the particle to be on the right side `slowly' increases. The remarkable result of our calculations is that the energy
scale over which the transition from symmetric ground state to completely asymmetric ground state can be made
arbitrarily small. As our `toy model' calculation demonstrated, this energy scale is controlled by the tunnelling probability
between the two wells, which is an energy scale that is not obviously present in the microscopic parameters (height and
width of the barrier). In fact, the better well-defined the two well system is, the smaller this energy scale. Fig.~3 (or Fig.~5)
demonstrates this quite dramatically, where imperceptibly small asymmetries in the potential give rise to a completely asymmetric
wave function.
There is very little indication of this failure from the ground state energy; instead, it
requires a calculation of the wave function to demonstrate this. 

These calculations serve to demonstrate a number of important principles for the novice. First, the numerical calculation is
`simpler' than the analytical calculation, and less likely to lead to error. By this we mean that solving for the wave function
through Eqs.~(\ref{allowed},\ref{relamp},\ref{relampinv}) is a little subtle and, for example, the wrong choice of using either
Eq.~(\ref{relamp}) or Eq.~(\ref{relampinv}) can lead to inaccuracies. In contrast the numerical solution is straightforward.
Then, the solution obtained here can be tied to perturbation theory. An accurate calculation of the energy can be achieved with
perturbation theory, but not so with the wave function! This ties into the variational principle, and teaches the important lesson that
a (very) accurate estimate of the energy certainly does not imply an even qualitatively correct wave function.

\begin{acknowledgments}

This work was supported in part by the Natural Sciences and Engineering Research Council of Canada (NSERC), by the Alberta
iCiNano program, and by a University of Alberta Teaching and Learning Enhancement Fund (TLEF) grant. 

\end{acknowledgments}

\end{document}